# Towards CW modelocked laser on chip - large mode area and NLI for stretched pulse mode locking

Neetesh Singh[1,2*], Erich Ippen[2], and Franz X. Kärtner[1]

[1]Centre for Free Electron Laser Science (CFEL)-DESY and University of Hamburg, Notkestrasse 85, 22607 Hamburg, Germany
[2]Research Laboratory of Electronics, Massachusetts Institute of Technology, 77 Massachusetts Avenue, Cambridge, MA 02139, USA
*neetesh.singh@cfel.de

**Abstract:**
Integrated modelocked lasers with high power are of utmost importance for next generation optical systems that can be field-deployable and mass produced. Here we study fully integrated modelocked laser designs that have the potential to generate ultrashort, high power, and high quality pulses. We explore a large mode area laser for high power pulse generation and study the various mode-locking regimes of dispersion managed soliton pulses in net anomalous and net normal dispersion cavities. Furthermore, we study numerically and experimentally general properties and tunability of a fast integrated saturable absorber based on low loss silicon nitride nonlinear interferometer. We believe this work guides the exploration of the future integrated high power modelocked lasers.

**Introduction:** Modelocked lasers (MLL) are a driving force behind many of the optical technologies such as nonlinear optical signal generation [1-4], optical frequency synthesis and broadband coherent communication [5-8], microwave signal generation [9], optical atomic clocks [10], photonic analog to digital conversion [11], astrocomb [12], volumetric 3D data storage [13], laser surgery [14], timing synchronization of high energy accelerator sources [15, 16], and many more. These are possible because a rare-earth gain based MLL produces ultrashort and high peak power optical pulses with low timing jitter [17, 18].

Passive MLLs have existed in fiber and bulk media for a very long time [19-26]. For the next generation systems, however, integration of the high power modelocked laser is important. Here, building up on previous demonstrations [27-30], we show proof of concept designs for integrated MLLs that can generate high peak power pulses with widths of 100fs and less, at high repetition rate (>1GHz). Although the lasers discussed here are at telecom wavelength using erbium doped aluminum oxide, they can easily be adapted to any other rare-earth materials. In the first section we discuss the design for a laser cavity with large lasing mode area (50μm$^2$) that facilitates increased power in the pulses as well as reduced optical nonlinearity-based pulse instability. Unlike the high power lasers, these designs support only large fundamental modes and thus avoid loss of power to higher order modes. These structures allow seamless transition of fundamental mode from a weak confinement to a high confinement region in a single layer thus allowing compact bends and dispersion management to leverage waveguide nonlinearity. In the second section we discuss pulse generation in net anomalous and net normal dispersion cavity using integrated saturable absorber based on a nonlinear interferometer (NLI) which is studied numerically and experimentally in the following section. The tunability of the reflection curve of an NLI using integrated heaters not only compensates for the fabrication tolerances, but more importantly gives an additional degree of freedom in modifying the saturable absorption strength, unlike a semiconductor saturable absorber. That can be useful for, self-starting, suppressing Q-switching instability, and varying the maximum allowed pulse power of an MLL (based on fibre, free space or integrated platform), which is normally fixed in a fiber or free space MLL based on nonlinear interferometer. Such an integrated NLI based saturable absorber will also find applications in Mid-IR modelocked lasers where reliable fast saturable absorbers do not exist [31], especially NLIs based on silicon, as it is transparent up to 8μm.

**Results:**
**1. Large mode area:** For high power fiber lasers it is customary to use large mode area fiber because it helps to reduce instability in the lasing mode by lowering the intensity and therefore the optical nonlinearity in the medium. A large mode area in the gain region also increases the saturation power to allow lasing with high power [32, 33]. For integrated lasers, not only a large mode area is required but seamless transition of the optical mode between the gain (weak confinement) and the guiding (tight confinement) section is desirable to leverage several functionality offered by integrated photonics. Such a device remains largely unexplored in integrated lasers [34-41]. In this proof of concept work we show waveguide designs on integrated platforms that support modes of area up to 50μm$^2$ for 1550nm, with >95% overlap between the pump and the signal, and offer tight mode confinement with single layer silicon nitride. Here we design the cavity for in-band pumping (1480nm), but 980nm can very well be implemented with slight modification of the waveguide dimensions. Moreover, these waveguides only support fundamental mode thus avoid loss of power to higher order modes that can plague fiber lasers, and therefore promise high power and high quality of the lasing mode [33].

Two different waveguide designs are investigated as shown in Fig.1. In Fig. 1a, the waveguide was designed with a silicon nitride (SiN) layer having the width (w) of 485 nm, height (h) of 800 nm and a silica buffer layer on top (g) of height of 450 nm. The aluminum oxide gain layer thickness is 2.2 μm. The transverse electric (TE) mode profile at 1480 nm (pump), which was calculated using a finite difference based mode solver (Lumerical), is shown in Fig. 1a. The effective area of the mode at the pump wavelength is ~ 50μm$^2$. The energy confined in the gain layer is ~ 98%. The pump and signal overlap is over 98%. The advantage of this waveguide design is that gain layer can either be directly deposited on the smooth top surface of the device directly out of the foundry, or can be bonded with a standalone gain medium, like a semiconductor gain medium, thus significantly lowering the propagation loss as it doesn't require any additional etching. Also this device does not require a relatively thicker (several microns) gain layer. Moreover, the modes can be pulled into the SiN layer very rapidly and can be fully confined in the SiN waveguide by increasing its width, thus offering transition between different sections, such as gain to saturable absorber or external cavity, without suffering from reflections due to index mismatch. Any spurious reflection becomes detrimental for the stability of the lasing modes thus making an MLL highly sensitive to these reflections [42, 43]. The mode can be switched from being fully in the gain medium to fully in the SiN layer by varying the SiN width to >45 nm (<100 nm for 1550 nm signal) which can be achieved adiabatically within a few microns length, as shown in Fig. 1(a1, a2 and a3). For shorter wavelengths the transition speed increases, for example for a waveguide designed to be pumped at 980 nm the width change required is merely 10-15 nm (not shown). Moreover, since the SiN layer is 800 nm thick, it allows anomalous dispersion by increasing the width of the SiN layer in the telecom window therefore supporting optical nonlinear functionality as well as dispersion compensation of a modelocked laser. Such a device does not require different vertical layers of SiN for mode transitions, tailored to different functions, which is prone to unwanted reflections due to optical impedance mismatch between different layers at the transition region [44-46], which is as mentioned above can be detrimental in a laser.

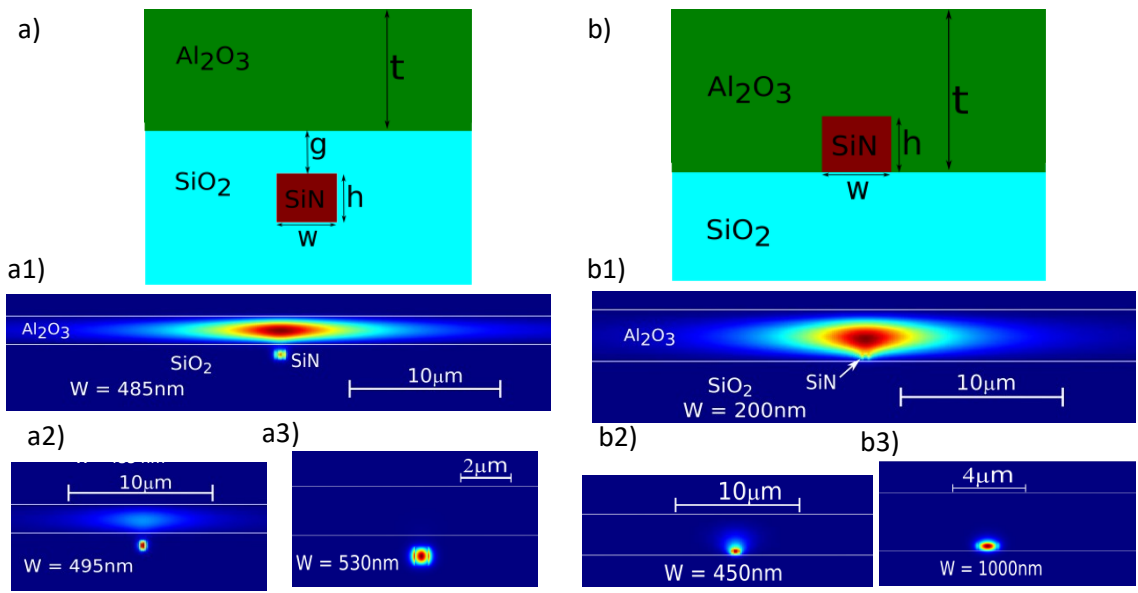

Fig.1 a) Large mode area waveguide with gain above a buffer layer of thickness *g*. a1) Mode profile of 1480 nm with 50μm$^2$ mode area. a2) and a3) modes with large SiN width demonstrating fast mode pulling. b) Large mode area with direct deposition of gain on SiN. b1) Mode profile at 1480 nm of area ~ 50μm$^2$. b2) and b3) Mode profiles with larger width of the SiN.

Another design investigated is shown in Fig.1 b, where there is no buffer layer and the gain is deposited directly on top of the SiN waveguide. The width of the SiN waveguide is 200nm, the thickness is 400nm and the aluminum oxide layer is 3.2 μm thick. The calculated mode area is ~50μm$^2$ for the pump and the overlap between the signal and the pump is over 95%. A similar device has been used previously although with mode area of roughly 5μm$^2$ [28]. The advantage of this device over the previous one (Fig.1a) is that it does not require a thick SiN layer, and requires less modification in the design for varying mode sizes, while also offering full signal mode confinement within the same SiN layer by increasing the width over 1000 nm. We must note that in this design we have not included the pedestal (which modifies the mode area), above the aluminum oxide layer which would be present after the deposition on top of the SiN waveguide, as it can be easily polished away. We also emphasize that in both of the designs even larger mode area can be obtained (~100μm$^2$), and due to tight confinement with increased SiN width, compact bends can be implemented in the laser to allow multiple straight gain sections.

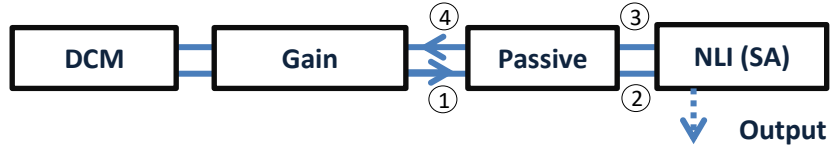

Fig.2 The modelocked laser simulation flow diagram. DCM is double chirped mirror for dispersion management. NLI is nonlinear interferometer, SA is saturable absorber. Numbers represent pulse monitoring regions.

**2. Modelocked laser:** In this section we present designs of integrated modelocked lasers based on net anomalous and net normal dispersion cavity [47-50]. We simulate a linear cavity due to the lack of a reliable isolator required for a ring cavity, which makes the system more sensitive to reflection and spatial hole burning but it has the advantage of ease in fabrication. In the simulation, we do not take into account spatial hole burning [51, 52] and reflection based instabilities [53].

In this model we solve for the pulse evolution in the cavity with different sections as shown in Fig. 2. Here the DCM (double chirped mirror) is an apodised chirped grating that provides the required dispersion [54-56], the gain is based on the rare earth material ($Al_2O_3:Er^{3+}$), the passive section is made of a low loss waveguide for attaining desired cavity length, and the NLI (nonlinear interferometer) is an intensity dependent interferometric reflector [57] acting as a saturable absorber (SA); more on the NLI in the later section. We note that the gain section here can replace the passive section with the gain adjusted such that the total roundtrip gain remains the same. The nonlinear Schrödinger equation (NLSE) for the pulse evolution in the cavity is given as:

$$\frac{\partial E}{\partial z} = \left[\frac{g-\alpha}{2} + \left(D_g - j\frac{\beta_2}{2}\right)\frac{\partial^2}{\partial t^2} + \frac{\beta_3}{6}\frac{\partial^3}{\partial t^3} - j\gamma(|E|^2)\right]E \quad (1)$$

Here $E$ is the pulse field envelop evolving along the length of the cavity. The NLI, a nonlinear Michelson type interferometer, is made of two arms attached to a coupler acting as a beam splitter with power split ratio of $r$ and $1-r$ in reflection and transmission, respectively, see section 3. We have used $r$ between 0.8 and 0.75. When the pulse enters the NLI the electric field in the high power arm is $E_h = \sqrt{r}E$, and in the low power arm is $E_l = i\sqrt{1-r}E$. When the pulse returns from the NLI back to the cavity the field envelope is given as, $E = \sqrt{r}E_h + i\sqrt{1-r}E_l$, and the signal out of the cavity from the NLI is given as, $E_o = i\sqrt{1-r}E_h + \sqrt{r}E_l$, which is the output of the laser in this configuration. Similarly, a nonlinear optical loop mirror (NOLM) or nonlinear amplifying loop mirror (NALM) can be implemented in the simulation if required. The gain $g = g_o/(1 + P_{avg}/P_{sat})$ where $g_o$ is the small signal net gain and the $P_{avg}$ is the average power in the cavity, and the saturation power $P_{sat} = E_{sat}/t_l$. The $t_l$ is the gain life time and the saturation energy is $E_{sat} = A_{eff}hf/e_c$, where $A_{eff}$ is the effective area of the lasing mode, $hf$ is the signal photon energy, and $e_c$ is the emission cross-sections. In the steady state when the gain is saturated, $P_{avg} = P_{sat}(g_o/l - 1)$, where $l$ is a function of the passive loss ($\alpha$), the dispersion of the gain ($D_g$) and the cavity ($\beta_n$) [47]. For a CW narrow linewidth laser $l$ would simply be $\alpha$. Usually for higher power lasers, high $P_{avg}$, that is high $P_{sat}$ and $g_o$, is required. As the net gain is limited by the amplified spontaneous emission (unless multi-stage gain is used), $P_{sat}$ can be increased by increasing the $A_{eff}$ while utilizing long gain life time which is required for low noise lasers. In the simulation, the gain, which is proportional to the pump power, is used such that the signal average power remains between 10 – 40 mW, which is similar to the experiment where over 50 mW intracavity power has been obtained [28]. The gain dispersion, $D_g = g/(2\pi\Omega)^2$, where $\Omega$ is the gain bandwidth (for $Al_2O_3:Er^{3+}$ it is 4 THz). The background loss $\alpha$ is 0.1 dB/cm. The nonlinearity, $\gamma = 2\pi(n_2)/\lambda A_{eff}$, where $n_2$ is the Kerr factor which is different for different sections of the cavity. The nonlinearity ($\gamma$) should be low for the gain and the passive section to avoid pulse instability [47, 58], that is because higher nonlinearity causes the competing weak pulses to have enough peak power to experience less loss through the SA. Unless large dispersion is used to spread the weak pulses in time they start to rob gain from the main pulse and eventually destabilize it after several roundtrips. For the simulation standard split-step Fourier method is used with a window size of 88ps and a grid size of 2.7fs. In all the simulations presented here, there was no noise added as the effect of the noise was negligible as was seen in the simulations (not shown). Also, although we have used a fixed gain bandwidth, larger gain bandwidth normally produces higher peak power and shorter pulses that stabilize relatively faster, that is with fewer roundtrips.

**2a. Net anomalous cavity:** Here we simulate a cavity with a negative net group delay dispersion (GDD), such as the one shown in Fig.3. In this we used the length of the gain section as 20 mm, passive section as 40 mm, NLI arms as 10mm, resulting in a repetition rate of 1.38 GHz. The $e_c$ is $8.3 \times 10^{-25} m^2$, with $t_l$ being 3ms. The $A_{eff}$ for the mode in the gain and passive is 50μm² and in the NLI it is 0.66μm². The NLI power split ratio is 80:20. The Kerr factor for the SiN = $3\times10^{-19}$m²/W, $Al_2O_3$ and $SiO_2$ = $3\times10^{-20}$m²/W. The dispersion in the gain is $\beta_{2g}$ = -$1.51\times10^{-26}$s²/m and $\beta_{3g}$ = $1.98\times10^{-40}$s³/m, in the passive it is $\beta_{2p}$ = $1.94\times10^{-25}$s²/m and $\beta_{3p}$ = $2.14\times10^{-39}$s³/m, and in the NLI it is $\beta_{2n}$ = $1.50\times10^{-24}$s²/m and $\beta_{3n}$ = $-5.33\times10^{-39}$s³/m. The dispersion of the DCM is adjusted to get the desired net anomalous dispersion. For the simulation the initial signal peak power was sub-femtowatt and the pulse (sech) width was 10ps. In Fig.3a we show the spectrum of stable pulses after 1000 roundtrips in a cavity with the net GDD of $-430\times10^3$fs². The average power of the stable pulses was 13 mW. We see the spectrum is similar at various locations in the cavity that is because the bandwidth of the pulse is narrow enough to avoid gain filtering. The Kelly sidebands appear, as expected for large net anomalous cavity, because the pulse experiences modulation of the nonlinear term $\gamma|E^2|$ (eq.1), through the cavity [59]. The peak power evolution of the pulse at different locations in the cavity is shown in Fig.3b. For the stable pulses the power is higher after the gain and drops slightly after the passive section due to the background loss and drops significantly in the NLI as the intensity dependent reflection is 40% at this peak power level (more in the NLI section), which is further reduced in the passive section. In Fig.3c the spectral evolution of the pulse is shown where the sidebands start appearing well before 100 iterations and reach full stability under 500 iterations, just like in the time domain shown in Fig.3d.

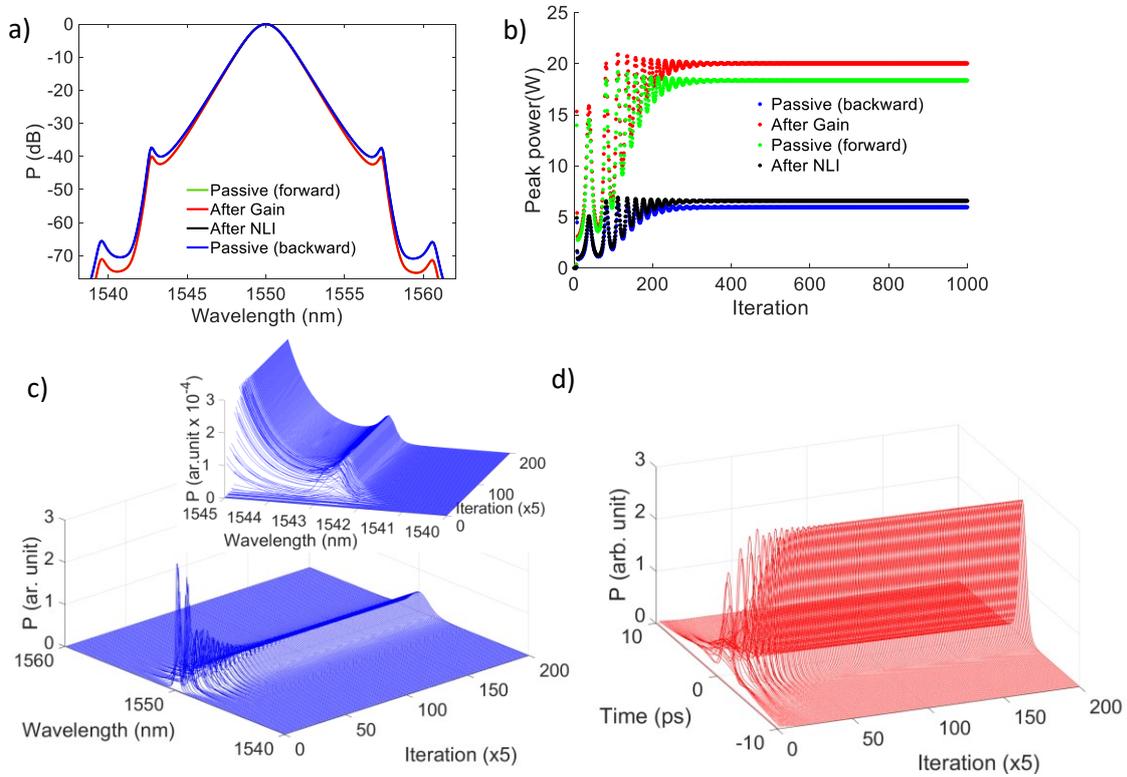

**Fig3.** Pulse evolution for the net GDD of -430kfs² (k is $10^3$). a) The pulse spectrum at various points in the cavity. Passive (forward) - after passive and before the NLI section (at 2 in Fig. 2), After gain - before the passive section (at 1 in Fig. 2), After NLI - before the passive section (at 3 in Fig. 2), and Passive (backward) - after passive and before the gain (at 4 in Fig. 2). b) The evolution of the peak power of the pulse. c) and d) The evolution of the pulse in the wavelength and time domain for 1000 iteration (After NLI).

Next, we show how the pulse evolves for varying net anomalous GDD. In Fig.4a the stable pulse spectra (After NLI, at position 3 in Fig. 2) for net GDD ranging from $-55\times10^3$ to $-455\times10^3$fs² is shown. Here the pump power thus the gain was adjusted for higher average power (around 40mW), which helps stabilizing the pulses for lower net anomalous GDD, therefore producing shorter pulses. As the net GDD is lowered the pulse starts getting broader in spectrum, the peak power increases, and also the Kelly sidebands start to weaken and eventually disappear near zero net GDD (see net normal cavity section). The pulse does not stabilize in this configuration at or below the net GDD of $-35\times10^3$fs² unless the pump power or the self-amplitude modulation (SAM) of the NLI, i.e. the strength of the saturable absorber, is increased. As mentioned before as the net GDD is reduced the susceptibility of the pulse to the weak competing long pulse (continuum) increases, because at lower net GDD the temporal broadening of the weak pulse is reduced causing increase in its peak power thus its higher reflectivity back to the cavity. By increasing the SAM, which is the rate of change of the reflectivity of the NLI with the peak power, see Section 3, the selectivity for a relatively higher power pulse is increased thus keeping the weak continuum from building up. For a cavity with a fixed saturable absorber, the pump power can also be increased to increase the overall initial pulse power so that the pulse reaches a relatively higher region of SAM of the NLI, which is strongest for a given saturable absorber between its minimum and the maximum reflection point and is zero at these points. Usually a lower SAM value of a saturable absorber in the Kerr based cavity contributes to non self-starting behavior. By mechanical vibration or active modulation technique in the cavity, the initial pulse gets enough peak power to be in the relatively higher SAM region thus helping it to self-start. The effect of the pump power and the SAM on the self-start issue has been studied in [60-62]. Additionally, the cavity configuration used in this section can be stabilized for a very low net GDD (~0fs$^2$) by increasing the SAM for the low power pulses just by applying an external phase bias on an NLI arm (using heaters, see NLI section),

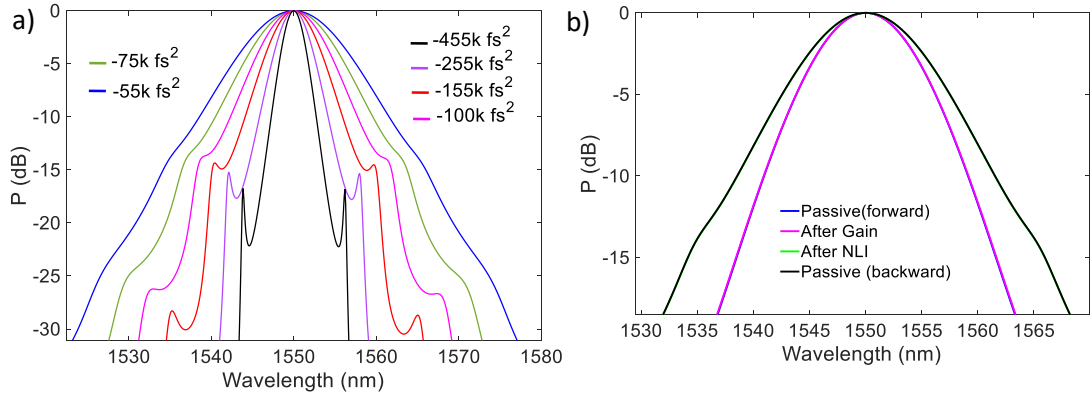

Fig.4 a) Pulse spectrum vs net GDD and b) spectrum for the pulse at different locations in the cavity with the net GDD of -55kfs$^2$ (k is 10$^3$). Passive (forward) - after passive and before the NLI section (at 2 in Fig. 2), After gain - before the passive section (at 1 in Fig. 2), After NLI - before the passive section (at 3 in Fig. 2), and Passive (backward) - after passive and before the gain (at 4 in Fig. 2)

In Fig.4b the spectra for the stable pulse is plotted for the net GDD of -55kfs$^2$ with peak power reaching in the gain section ~ 100W. Here we see the spectrum changes significantly at different locations of the cavity. That is because the gain filtering is strong for such a broadband pulse. The pulse regains its bandwidth in the NLI which has a strong self-phase modulation due to silicon nitride's high Kerr factor. In Fig 5 we plot the output pulse from the NLI (reject port), in time (Fig. 5a) and wavelength (Fig.5b) for the net GDD of -55kfs$^2$ having peak power of 20W and the pulse width of < 1ps which is chirped. The output pulse shape is flatter, in fact having weak shoulders, whereas the pulse back to the cavity is narrower in time. That is because the output is from the rejection port of the NLI which gets the rejected wings of the main pulse thus making the cavity pulse shorter in time, see section 3. Pulses with shape like the one in the cavity can be extracted for the output with an additional drop port anywhere in the cavity albeit with an additional loss.

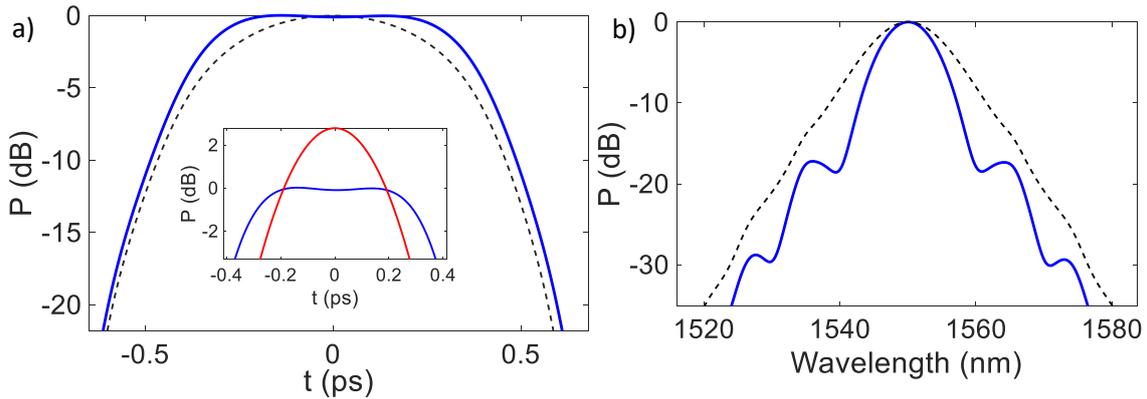

Fig.5 a) The output pulse in time domain from the NLI reject port (solid blue) and the pulse back to the cavity after the NLI (dash). The compressed output pulse (red) with an smf28 fiber (inset). b) The spectrum for (a). The net GDD of the cavity is -55kfs$^2$.

The cavity pulse energy is 20pJ and the output pulse energy is roughly 10pJ, the peak power of which can be increased by a factor of 2 with external compression using a waveguide or a fiber, to reach 40W with pulse width of 400fs.

**2b. Net normal cavity:** In this section the net GDD of the cavity was kept normal. For the simulation we have used similar parameters as in the previous section except the length of the NLI is 22 mm resulting into repetition rate of 1.15 GHz, and the NLI power split ratio is 75:25, and as before the DCM dispersion is varied to obtain net normal GDD. In Fig. 6a we have the spectrum of the stable pulse (after 1000 iteration) at various points in the cavity of net GDD of $110 \times 10^3$ fs$^2$. The pulse has some spectral filtering after the gain and gets broad again after the NLI. The shoulders appear in the pulse spectrum as in an all normal dispersion laser [26], and in fibers with self-phase modulation [62, 63]. The peak power evolution of the pulse at different locations in the cavity is shown in Fig.6b, which follows similar trend to that in Fig. 3b. The pulse stabilizes faster than with the net negative GDD (Fig. 3), because the average power in the cavity here is higher, around 24 mW.

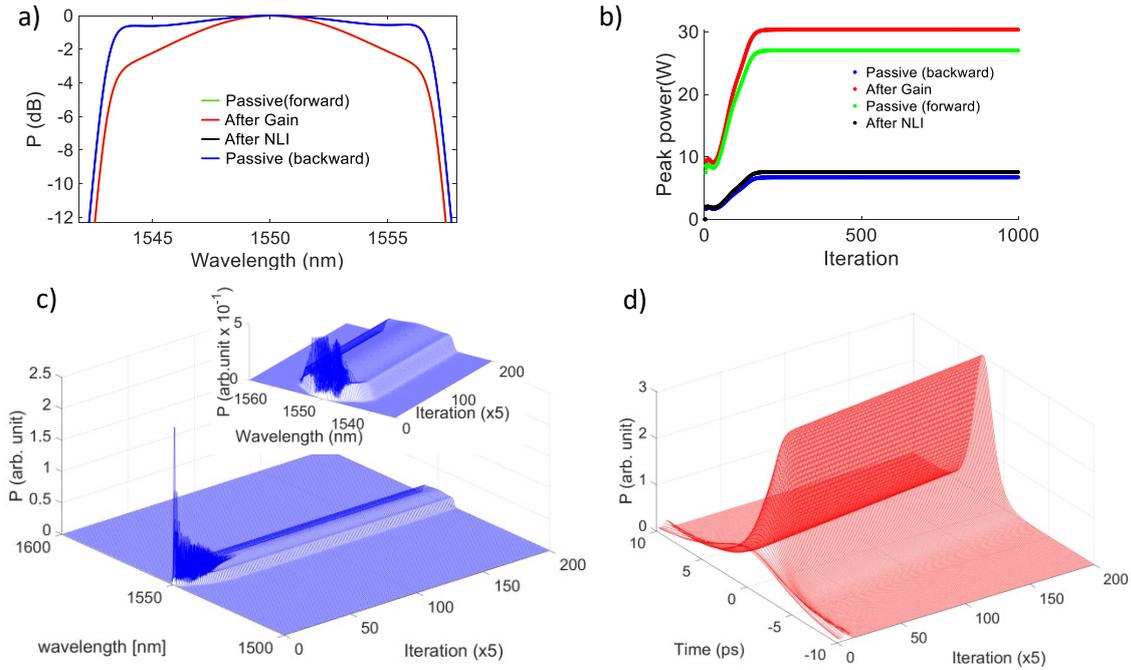

**Fig6.** Pulse evolution for the net GDD of +110kfs2 (k is $10^3$). a) The pulse spectrum at various points in the cavity. Passive (forward) - after passive and before the NLI section (at 2 in Fig. 2), After gain - before the passive section (at 1 in Fig. 2), After NLI - before the passive section (at 3 in Fig. 2), and Passive (backward) - after passive and before the gain (at 4 in Fig. 2). b) The evolution of the peak power of the pulse. c) and d) The evolution of the pulse in the wavelength and time domain for 1000 iteration (After NLI).

We used higher power and stronger NLI here because the pulse tends to have lower peak power in the normal dispersion NLI making it harder to modelock the pulses. The spectral and temporal evolution are shown in Fig. 6c and d, where we see the pulse reaching stability after close to 200 iterations. In Fig. 7a we show the pulse spectra (After NLI, at position 3 in Fig. 2) for different net normal GDD of the cavity with the average power of ~24 mW. As the net GDD gets more positive the pulse starts getting narrower, the peak power reduces, and the shoulder gets stronger. In this configuration to get stable pulses above 110kfs$^2$ one needs to either increase the pump power or the SAM of the NLI. With lower net GDD as in the net anomalous case the pulse gets broadband and shorter in time.

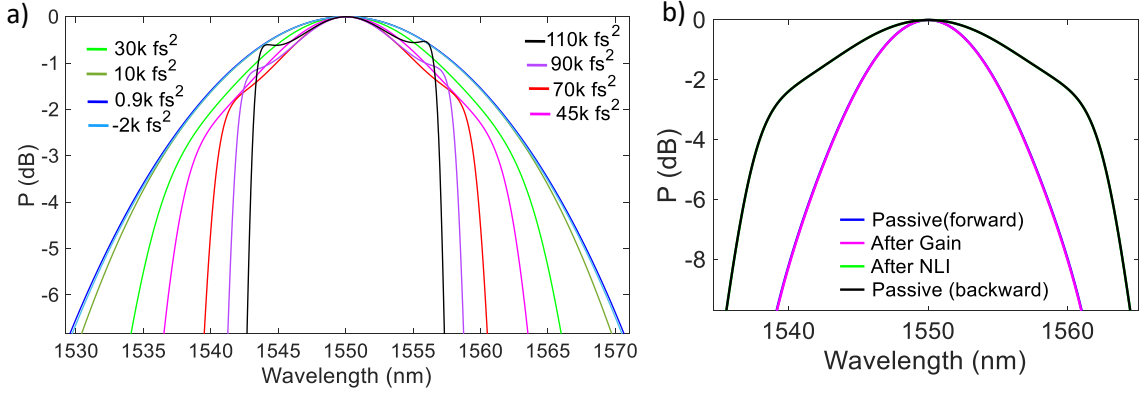

**Fig.7 a)** Pulse spectrum vs net GDD and b) spectrum for the pulse at different location in the cavity with the net GDD of +45kfs$^2$ (k is 10$^3$). Passive (forward) - after passive and before the NLI section (at 2 in Fig. 2), After gain - before the passive section (at 1 in Fig. 2), After NLI - before the passive section (at 3 in Fig. 2), and Passive (backward) - after passive and before the gain (at 4 in Fig. 2).

Moreover, this device can work in a cavity with dispersion ranging from net normal to net anomalous, as can be seen in Fig.7 a, where the pulse is stable in the negative (-2kfs$^2$) close to zero net cavity dispersion. That is because the NLI has higher SAM than in the previous section. The highest intracavity peak power for the net GDD of 0.9kfs$^2$ was 150W after the gain section. In Fig. 7b we show large spectral variation due to the gain filtering (for net GDD +45kfs$^2$) with roughly 10-15 nm bandwidth narrowing in the gain section. The pulse recovers in bandwidth after the NLI, however due to high chirp in the cavity the peak power remains low. Also, as seen in Fig.8a, the pulses from the NLI and back to cavity look very similar. That is because the peak power of the pulse in the NLI is low due to high chirp, and therefore the pulse shortening effect of the NLI is not strong. The cavity pulse energy is 20pJ and the output pulse energy is roughly 10pJ, i.e. the output peak power is 8W. The output pulse can be compressed externally up to 90W with an almost 10x compression, producing pulses as short as 120fs [49].

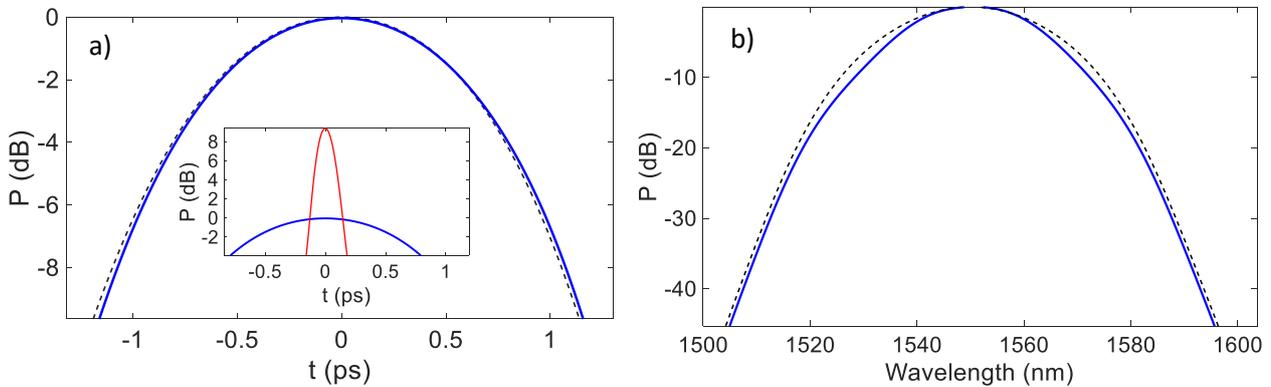

Fig.8 a) The spectrum of the output pulse from the NLI reject port (solid blue) and the pulse back to cavity after the NLI (dash). The compressed output pulse (red) with an smf28 fibre (inset). b) The spectrum for (a). The net GDD of the cavity is 0.9kfs$^2$.

In the net normal and the net anomalous GDD cavity simulations, to keep the computation time short we used low pump power to keep the avergae power within 40mW. In a real device this can be higher, thus one can achive higher peak power pulses with shorter pulse widths. Furthermore, the cavity length can also be increased to increase the power in each pulse. And as is routinely done, an external gain section can also be used to amplify pulses, for example with an integrated gain waveguide that can provide high net gain >10 dB [65-68]

**3. Nonlinear interferometer (NLI) as a saturable absorber:** A modelocked laser requires a device that provides higher gain for shorter pulses than for longer pulses. This can be achived with an intensity sensitive reflector (a saturable absorber) inside the cavity as short pulses tend to have high peak power. An NLI is such a device based on Kerr nonlinearity and interference. The idea of shortening the pulse with an interferometric system has been employed in the last few decades using external nonlinear Fabry-Perot interferometers and nonlinear Michelson interferometers [57, 69-73]. Kerr based saturable absorbers have been in use for many years with different modelocking schemes. For example Kerr lens modelocking (KLM), based on self focusing, has been used to generate pulses as short as 5fs [25]. Additive pulse mode locking [71, 72] relying on a Fabry-Perot based external nonlinear cavity for pulse shortening made modelocking possible with various fiber based systems and was extended into many other interfereometric versions based on nonlinear polarisation rotation (NPR) [74], a nonlinear optical loop mirror (NOLM) [57] and a nonlinear amplyfying loop mirror (NALM) [75]. A Kerr based SA has the advantage over a semiconductor based SA in that it tends to produce shorter pulses, has a high damage threshold and is broadband. However it usually has self starting issues, mainly due to the low SAM at low power when the pulse is building up as was observed before[62]. In this work we utlize silicon nitride waveguide for a nonlinear Michelson type saturable absorber that can be easily integrated [76], and due to its high nonlineairty compared to a silica fibre it can potentially facilitate self starting. Additionally, the relative ease in tunning the reflectivity of an NLI through externally applied phase bias (see below), the SAM can be increased which can be useful to help self-start an MLL, and avoding Q-switching instability for lower pump powers.

In this work we have used a Michelson type nonlinear interferometer as shown in Fig 9a (inset). It is based on a coupler with $r > 0.5$, so that one arm has higher power than the other for obtaining a differential power depenendent phase shift. Maximum reflection back to the cavity happens when the differential phase shift, $\Delta\Phi_n + \Delta\Theta = m\pi$, where $\Delta\Phi_n$ is the nonlinear phase shift difference between the high power arm ($r$) and the low power arm ($1-r$), and the $\Delta\Theta$ is any additonal phase shift difference due to fabrication tolerance or externally applied bias (as is discussed later), and $m$ is an odd integer (for minimum reflection $m$ is an even integer). The reflection strength cycles through its maxima and minima with the pulse peak power. If $r = 0.5$, the $\Delta\Phi_n$ will be zero thus no light will be reflected (assuming $\Delta\Theta=0$) . To understand the key dynamics of an NLI we have calculated pulse evolution using NLSE, as in Eq.1 (where the gain is zero). The reflection curve (Reference) shown in Fig.9a was calculated with a 300fs pulse into the NLI arms of 4 mm length, $r$ =0.8, effective area of the mode = $0.85\mu m^2$, waveguide loss = 0.1dB/cm, Kerr factor = $3\times10^{-19}m^2/W$. With the increase in power the reflectivity back to cavity increases until the differential phase shift reaches $\pi$. The upward slope, upto the 1$^{st}$ peak, is the region where an NLI is used as a saturable absorber, in the downward slope it is used as a pulse limiter to stabilize pulses [77].

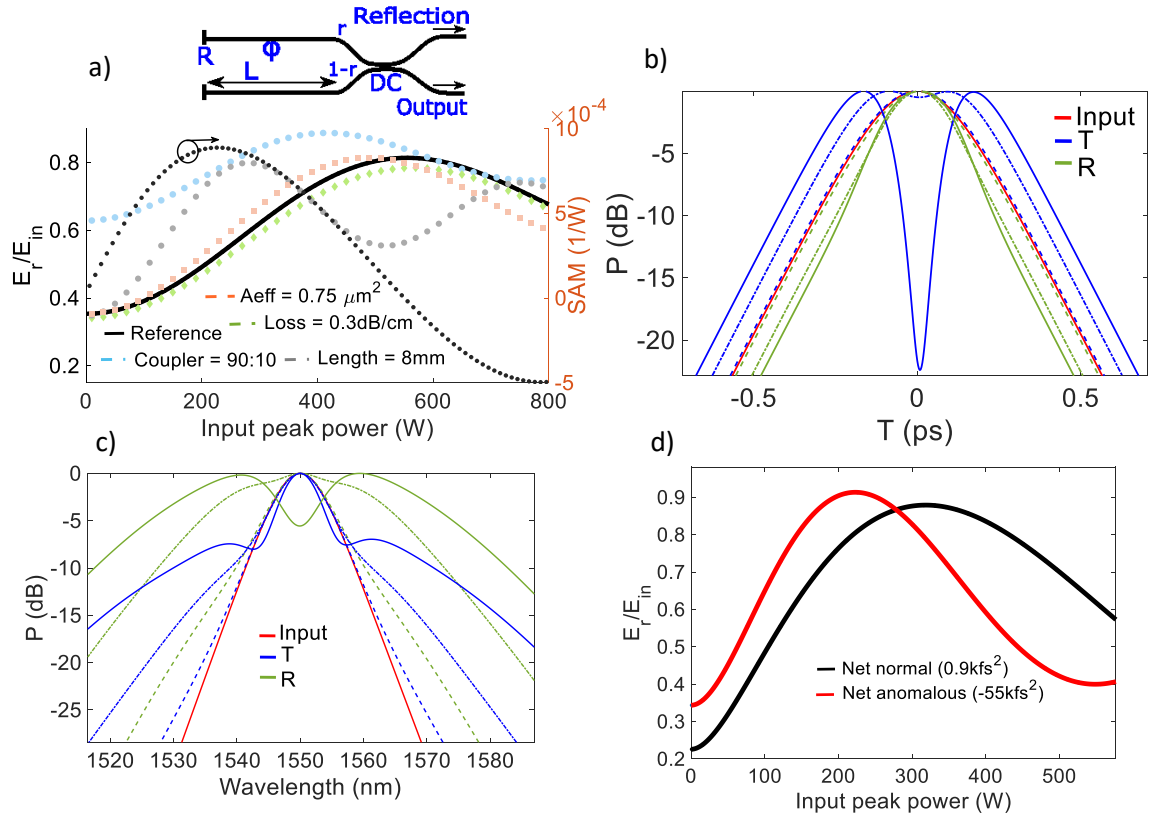

Fig.9 a) The NLI reflection curves for different device parameters. Schematic of the NLI (inset), where reflection is the signal back to cavity, Φ is the nonlinear phase shift, L is the length of the cavity, r is the coupler power splitting factor, R is the reflectivty of each waveguide which is 100% in the simulation. Er and Ei are the energy of the reflected and the input pulse. The SAM curve (black dots) is shown for the reflection curve (black solid). b) The pulse in time domain for different powers. T is the output from the NLI, R is the reflection. Dash, dash-dot and solid curves are for 80W, 250 W, 450 W respectivley. c) The spectrum for the pulses in b). d) The NLI curve for the MLL simulated in the section 2a and 2b for a cavity with net GDD = -55kfs$^2$ and 0.9kfs$^2$, respectivly.

Self-amplitude moduoltion (SAM) is the slope of the reflection curve, as shown in Fig.9a, which determines the strength of an NLI as a saturable absorber. SAM = d$R$/d$P$(1/2√$R$), where $R$ is $E_r$/$E_{in}$, i.e the ratio of the energy of the reflected and the input pulse, and $P$ is the input peak power. SAM determines how strongly a pulse of certain peak power will be reflected, hence experience lower loss in the cavity than the weak competing long pulses. SAM is zero at the maxima and the minima of a reflection curve and peaks where the slope is the strongest. The higher the SAM the stronger the reflectivity of a pulse for a given power compared to background weak pulses, thus a larger SAM helps self-starting an MLL. However, as the SAM increases with the pulse power, the laser can go into Q-switching instability [62, 78], which can be overcome by increasing the pump power to help raise the pulse peak power to reach the lower SAM region near the 1$^{st}$ reflection maxima, see Fig.9a. Also the limit on the maximum power achievable, determined by the 1$^{st}$ reflection peak of an NLI, can be varied to a high or a low power with the help of tunability of the reflection curve by varying external phase bias (ΔΘ) (see below). In the simulation presented in this section we have not included dispersion. With normal dispersion waveguides in an NLI the peak of the reflectivity curve shifts to higher power as the pulse peak power is reduced, and hence the overall phase-shift is reduced. Anomalous dispersion in an NLI can cause opposite effects along with the soliton effect which has been used for soliton switching [79]. The general trend, however, can be extracted by only including the nonlinear term.

Next we show the dependence of the NLI's reflectivity curve on its various parameters. By reducing the effective area ($A_{eff}$) the intensity in the NLI is increased thus $\Delta\Phi_n$ is increased casuing the peak to shift to lower power compared to the reference. With higher loss (0.3dB/cm) the overall curve shifts down (due to high propagation loss) and the peak shifts to higher power (by 15W) as more power is required for π shift. Increasing the length of the NLI affords larger $\Delta\Phi_n$ thus the peak shifts to lower power, but due to higher total propagation loss the peak drops by 0.015% compared to the reference. Also we observe the reflectivity of the 2$^{nd}$ minima and the maxima are different from the 1$^{st}$. That is because with high peak power the nonlinear phase shift range is large across the width of the pulse, and hence only part of the pulse is transmitted or reflected, which gets even more complicated with dispersion. By increasing the $r$ the reflection peak shifts to lower power as there is more power now in the high power arm giving high $\Delta\Phi_n$, this however reduces the SAM. In Fig.9 b we show how the pulse at different peak powers evolves through the NLI (Reference in Fig. 9a) in single pass. Signals are normalised to their maximum to clearly see the changes in shape. At low power, with an 80 W pulse (dash curve), the $\Delta\Phi_n$ is small; thus there is weak pulse shaping effect and all the signal, reflected and the output has very little variations. As the power is increased to 250W (dash-dot), there is strong $\Delta\Phi_n$ at the peak of the pulse compared to the sides, hence a large part of the pulse is reflected, and the sides are rejected through the output (a dip at the centre can be seen in the output pulse). This helps shortening the pulse back to the cavity. As the power is increased further, 450W (solid), a larger part of the pulse acquires nonlinear phase shift, hence causing sronger reflection leaving a larger dip in the output signal. Here, since we are not using dispersion, pulse shortening is manily due to the nonlinear interference in the NLI and not due to the self-phase modulation in the high power arm of the NLI. The corresponding spectra are shown in Fig. 9c.

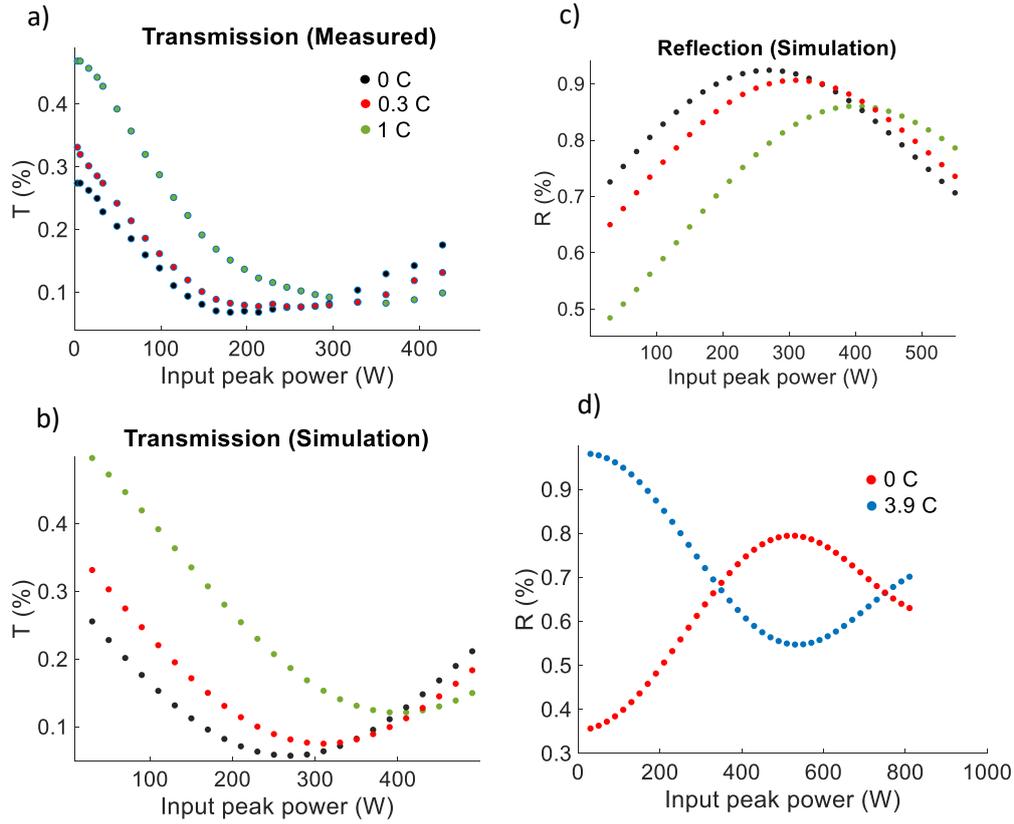

Fig10. a) Measured transmission of an NLI for different thermal phase shift. b) and c) the simulated transmission and reflection curve, respectivley. d) Reflection curve for an NLI without thermal or fabrication tolerance phase phase shift (0°C), and with thermal phase shift of π (3.9°C).

In Fig.9d we show the NLI reflection curves for the MLL with the net anomalous and net normal disperion of -55kfs$^2$ and 0.9kfs$^2$, studied above in section 2a and 2b. The curves are not only different due to the differences in the NLI's length and the coupler power splitting ratio (*r*), but also due to the difference in the pulses entering the NLIs. The difference in the pulse shaping between the net anomalous and the normal cavity seen in Fig.5a and Fig.8a is due to the fact that the pulse in the NLI, in the net anomalous dispersion cavity, has relatively larger nonlinear phase shift (similar to the high power case in Fig.9c) compared to the one in the net normal cavity (similar to the low power case in Fig.9c). To increase the pulse shaping in the net normal cavity and also faciliate pulse stablilty at lower power, a lower or anomalous dispersion of the NLI in the net normal cavity can be used while making sure not too high of anomalous dispersion in the NLI is used as that can cause the competing weak pulse to acquire high enough peak power to destabelise the modelocking.

**3a NLI Measurement:** We have designed and tested an NLI fabricated with silicon nitride waveguide using low pressure chemical vapour deposition (LPCVD) technique. To account for the fabrication tolerance and applying external phase shift bias ΔΘ (see above), heater layers were deposited on top of the waveguides. The length of the waveguides in the NLI is 4mm, $r$ = 0.8, $A_{eff}$ = 0.85μm$^2$, the dispersion $β_2$ = -1.25x10$^{-25}$ s$^2$/m and $β_3$ = -8.15x10$^{-40}$s$^3$/m, and the waveguide width = 1.1μm and the thickness = 800nm. The reflectors at the end of the waveguides were made of loop mirrors. The device was tested with a laser having 200fs pulses at 80 MHz repetition rate. The output was collected from the NLI's ouput (rejection port), as shown in Fig 10a with the data taken for different heater temperatures. To obtain 0.3°C and 1°C temprature changes, corrosponding to 0.08π and 0.26π phase shifts, 6mA and 10 mA currents were applied, respectively. The simulations of the transmission and reflection curves are shown in Fig.10(b & c) that also suggest that the fabrication tolerance based phase difference between the two waveguides of the NLI was ~ 4π which can happen with micron level length variation. The peak of the curve moves to high power, see reflection curves, by applying thermal phase shift on the high power arm. We also observe the peak of the reflection is dropping in magnitude with the thermal phase shift. This can be understood by seeing the Fig. 10d. Here the device was simulated without any external or fabrication tolerance based phase shift with 0°C (an ideal condition). By appling a phase shift of π with 3.9C (on the low power arm) the curve's peak shifts to zero power and also the magnitude of the reflection increases to almost 100%, compared to the 0°C case of 80%. That is because, near zero power there is almost negligble nonlinear phase shift ($\Delta\Phi_n$) across the pulse from the center to the edges. By applying a constant π phase shift, the total differential phase shift at every point in the pulse becomes almost a π, therefore the entire pulse gets reflected with of course without any pulse shaping. However, as the peak is shifted to higher power, the pulse does not have neglilgble phase shift across its width anymore i.e. the peak has significantly higher phase shift than the sides. Therefore, a constant phase shift causes some part of the pulse to be of π and some part to be of different phase, hence only part of the pulse gets refelected back. That is the reason for the reduction in magnitude of the peak of the reflection curve as it shifts to higher power, seen in Fig. 10c. This suggests that low power pulses of any shape will be reflected more effciently than high power pulses, which can find applcations where close to full reflection/switching is required for low power signal.

Furthermore, the peak of SAM of an NLI can be shifted to be near zero pulse power by applying hearter on the low power arm as that shifts the 1$^{st}$ reflection peak towards lower power. For example, in Fig. 10d the reflection curve for 0ºC can be moved to the left towards low power such that the slope of the curve (SAM) is the strongest near zero pulse peak power. This will be useful, as a higher SAM helps self starting and stabelising the pulses, especially for a cavity having high nonlinearity and low net GDD, without requiring to modify the length and the coupling ratio ($r$) of the NLI. Moreover, once the pulse has stablised near the 1$^{st}$ reflection peak, by applying thermal phase shift on the high power arm the peak of the reflection curve can be shifted to higher power thus increasing the maximum achievable pulse power limit.

In conclusion, we have shown designs of large mode area waveguides for integrated modelocked lasers that do not support higher order modes unlike the fibre lasers. We also simulated net normal and net anomalous modelocked laser cavities showing different pulse dynamics similar to what has been seen in the experiments with fiber lasers [48], showing potential for generating high output power and short pulses without any additional amplifier. Furthermore, we studied an integrated nonlinear Michelson interferometer as a fast saturable absorber numerically and experimentally. We believe this work will help the design of the future high power modelocked lasers on chip.

**Acknowledgment:** This work was supported by the DFG Priority Program SP2111 under contract PACE.

**Competing interests:** The authors declare no competing interests.

**References:**
1. J. K. Ranka *et. al.* Visible continuum generation in air-silica microstructure optical fibers with anomalous dispersion at 800 nm. *Opt. Lett.* **25**, 25-27 (2000).
2. Alfano RR, Shapiro SL . Observation of self-phase modulation and small-scale filaments in crystals and glasses. *Phys Rev Lett* 1970; **24**: 592–594.
3. N. Singh, *et. al.* Octave-spanning coherent supercontinuum generation in silicon on insulator from 1.06 μm to beyond 2.4 μm. *Light Sci. Appl.* **7**, 1-8, (2018).
4. J. Leuthold, Nonlinear silicon photonics. *Nature photon.* **4**, 535-544 (2010).
5. J. L. Hall, Nobel Lecture: Defining and measuring optical frequencies. *Rev. Mod. Phys.* **78**, 1279–1295 (2006).
6. N. Singh, M. Xin, N. Li, D. Vermeulen, A Ruocco, E. S. Magden, K. Shtyrkova, P. T. Callahan, C. Baiocco, E. Ippen. F. X. Kaertner, M. R. Watts, "Silicon photonics optical frequency synthesizer-SPOFS," CLEO ATh4I.2 (2019).
7. T. R. Schibli, K. Minoshima, F.-L. Hong, H. Inaba, Y. Bitou, A. Onae, H. Matsumoto, "Phase-locked widely tunable optical single-frequency generator based on a femtosecond comb," Opt. Lett., 30, 2323-5 (2005).
8. M. Nakazawa, K. Kikuchi, T. Miyazaki, *Spectral density optical communication* (Springer, 2010)
9. T. M. Fortier, M. S. Kirchner, F. Quinlan, J. Taylor, J. Bergquist, T. Rosenband, N. Lemke, A. Ludlow, Y. Jiang, C. Oates, et al., "Generation of ultrastable microwaves via optical frequency division," Nat. Photonics, 5 (7), p. 425 (2011).
10. S. A. Diddams *et al.*, "An optical clock based on a single trapped $^{199}$Hg$^{+}$ ion", Science 293, 825 (2001).
11. A. Khilo, et. al. , "Photonic ADC: overcoming the bottleneck of electronic jitter," Opt. Express 20:(4) pp. 4454-4469 (2012).
12. Wilken, T. et al. A spectrograph for exoplanet observations calibrated at the centimetre-per-second level. Nature 485, 611–614 (2012).
13. D. A. Parthenopoulos et.al, "Three dimensional optical storage memory". Science, 25 (1989).
14. T. Juhasz, et.al "Corneal Refractive Surgery with Femtosecond Lasers", J. Sel. To. Quant. Elec. 5 (1999).
15. A. L. Cavalieri*, et. al.* Clocking femtosecond X rays. *Phys. Rev. Lett.* 94, 114801 (2005).
16. J. Kim, J. A. Cox, J. Chen, F. X. Kärtner. Drift-free femtosecond timing synchronization of remote optical and microwave sources. *Nat. Photon*. **2**, 733-736 (2008).
17. J. Kim and F. X. Kärtner, "Attosecond precision ultrafast photonics," Laser & Photon. Rev., 1–25 (2009).
18. H.A. Haus, A. Mecozzi, "Noise in mode-locked lasers", IEEE J. Quant. Elect. QE-29, 983 (1993).
19. A. J. DeMaria, D. A. Stetser, and H. Heynau, Self modelocking of lasers with saturable absorbers," Appl. Phys. Lett. 8(7), (1966).
20. E. Ippen and C. Shank, Sub-picosecond, kilowatt pulses from a mode-locked CW dye laser," IEEE J. Quan. Elec. 10(9), 1974.
21. U. Keller, et. al Semiconductor saturable absorber mirrors (SESAM's) for femtosecond to nanosecond pulse generation in solid-state lasers," IEEE J.Sel.to.Quant.Elec. 1996.
22. D. E. Spence, P. N. Kean, and W. Sibbett, 60-fsec pulse generation from a self-mode-locked Ti:sapphire laser," Optics Letters 16, 1991.
23. T. Brabec, C. Spielmann, P. F. Curley, and F. Krausz, \Kerr lens mode locking," Opt. Letters 17, 1992.
24. E. P. Ippen, H. A. Haus, and L. Y. Liu, Additive pulse mode locking," JOSA, B 6(9)1989.
25. R. Ell et. al. Generation of 5-fs pulses and octave-spanning spectra directly from a Ti:sapphire laser," Opt. Lett. 26, 2001.
26. A. Chong et. al. "All-normal-dispersion femtosecond fiber laser", Opt. Express, 21 (2006).
27. F. X. Kärtner, *et. al.* Integrated rare-Earth doped modelocked lasers on a CMOS platform. *SPIE* (2018).

28. K. Shtyrkova, et. al. CMOS-compatible Q-switched mode-locked lasers at 1900nm with an on-chip artificial saturable absorber. *Opt. Express.* **27**, 3542-3556 (2019).
29. K. Shtyrkova "Fully integrated CMOS-compatible mode-locked lasers," *Thesis* MIT (2018).
30. H. Byun, et. al., ""Integrated low jitter 400-Mhz femtosecond waveguide laser, IEEE Phot. Tech. Lett, 21, 2009.
31. J. Ma, et. al. "Review of mid-nfrared mode-locked laser sources in the 2um-3um spectral region", Appl. Phys. Rev. 6, 021317 (2019).
32. J. Nilsson, et. al "Modeling and optimization of low-repetition-rate high-energy pulse amplification in cw-pumped erbium-doped fiber amplifiers", Opt. Lett. 18, 1993
33. D. J. Richardson et. al. "High power fiber lasers: current status and future perspectives" JOSA B, 27 (2010).
34. E. H. Bernhardi, et. al., "Ultra-narrow-linewidth, single-frequency distributed feedback waveguide laser in Al2O3:Er3+ on silicon", Opt. Lett. **35**, 2394 (2010).
35. J. D. B. Bradley, et. al., "Integrated Al2O3:Er3+ ring lasers on silicon with wide wavelength selectivity", Opt. Lett. **35**, 73 (2010).
36. P. Purnawirman et. al. "C- and L-band erbium-doped waveguide lasers with wafer-scale silicon nitride cavities, Opt. Lett. 38, 2013.
37. E. S. Hosseini et. al.,"CMOS-compatible 75 mW erbium-doped distributed feedback laser", Opt. Lett. 39, 2014.
38. E. S. Magden, et.al.,"Monolithically-integrated distributed feedback laser compatible with CMOS processing." Opt. Express, 25 (2017).
39. N. Li et. al., "Monolithically integrated erbium-doped tunable laser on a CMOS-compatible silicon photonics platforms" Opt. Express 26, 2018.
40. J. P. Donnelly, et. al. "AlGaAs-InGaAs slab coupled optical waveguide lasers" IEEE. J. Quant. Electr., 39, 2 (2003).
41. A. Y. Liu, et. al. "Quantum dot lasers for silicon photonics" Photon. Research, 3 (2015).
42. H. A. Haus, and E. P. Ippen "Self-starting of passively mode-locked lasers" Opt. Lett. 16, 1991.
43. K. Tamura, et.al. "Self-starting additive pulse mode-locked erbium fibre ring laser." Elec. Lett. 28, 1992.
44. R. Sun, et. al. "Impedance matching vertical optical waveguide couplers for dense high index contrast circuits", Opt. Express. 16, (2008).
45. D. D. John et. al. "Multilayer platform for ultra-low-loss waveguide applications", IEEE. Photon. Tech. Lett. 24, 11, (2012).
46. W. D. Sacher, et. al. "Tri-layer silicon nitride-on-silicon photonic platform for ultra-low-loss crossings and interlayer transitions", Opt. Express. 25, 25, (2017).
47. H. A. Haus, J. G. Fujimoto, and E. P. Ippen "Structures for additive pulse mode locking" J. Opt. Soc. Am. B, 8, 1991.
48. K. Tamura, et. al. "Soliton versus nonsoliton operation of fiber ring lasers", Appl. Phys. Lett., 64 (1994).
49. K. Tamura, et. al., "77-fs pulse generation from a stretched pulse mode locked all fiber ring laser", Opt. Lett. 18, 1993.
50. H. A. Haus, et. al. ,, Stretched-pulse additive pulse mode-locking in Fiber ring lasers: theory and experiments" IEEE J. Quant. Elec., 31, 1995.
51. B. Braun, et. al.," Continuous-wave mode-locked solid state lasers with enhanced spatial hole burning", Appl. Phys. 61, 429 (1995)
52. F. X. Kartner, et.al.,"Continuous-wave mode-locked solid state lasers with enhanced spatial hole burning" Appl. Phys, 61, 1995.
53. M. Homar et. al. "Travelling wave model of a multimode Fabry-Perot laser in free running and external cavity configuration ", IEEE. J Quant. Electron. 32, 1996.
54. F. X. Kartner, N. Matuschek, T. Schibli, U. Keller, H. A. Haus, C. Heine, R. Morf, V. Scheuer, M. Tilsch, and T. Tschudi, ''Design and fabrication of double-chirped mirrors,'' Opt. Lett. 22, 831–833 (1997)
55. F. Ouellette "Dispersion cancellation using linearly chirped Bragg grating filters in optical waveguides" Opt. Lett. 12, (1987).
56. P. T. Callahan, et.al.,"Double chirped Bragg gratings in a silicon nitride waveguide." CLEO, SF1E (2016).
57. N. J. Doran and D. Wood, "Nonlinear optical loop mirror", Opt. Lett. 13, 1988.
58. I. D. Jung, et. al.,"Experimental verification of soliton mode locking using only a slow saturable absorber." Opt. let. 20, 1995.
59. S. M. Kelly, "Characteristic sideband instability of periodically amplified average soliton", Electron. Lett. 28, 1992.
60. E. P. Ippen, et.al.,"Self-starting condition for additive pulse mode locked lasers." Opt. Lett. 15, 1990.
61. F. Krausz, et. al. "Self starting passive modelocking". Opt. Lett. 16, 1991.
62. C. Spielmann et.al. "Experimental study of additive pulse mode locking in an Nd:Glass laser." IEEE J. Quant. Elect. 27, 1991.
63. G. P. Agrawal "Nonlinear fiber optics" 5th ed. Chpt 4. 2013.
64. A. M. Johnson, and W. M. simpson "Tunable femtosecond dye laser synchronously pumped by the compressed second harmonic of Nd:YAG", J. Op. Soc. Am. B. 4, 1985.
65. J. D. Bradley, et.al. "Erbium doped integrated waveguide amplifiers and lasers." Laser Photonics Rev. 5, (2011).
66. J. Shmulovich et. al. "Integrated planar waveguide amplifier with 15 dB net gain at 1550 nm" OFC, 1999.
67. J. Ronn, et.al. "Ultra-high on-chip optical gain in erbium-based hybrid slot waveguides" Nat. Comm. 10, 2019.
68. S. A. Vazques-Cordova, et.al. "Erbium-doped spiral amplifiers with 20dB 0of net gain on silicon" Opt. Express, 22 (2014).
69. T. N. C. Venkatesan et.al.,"Optical pulse tailoring and termination by self-sweeping of a Fabry-Perot cavity." Opt. Commun, 31, (1979).
70. H. J. Eichler, et.al.,"Modulation and compression of Nd:Yag laser pulses by self tuning of a silicon cavity." Opt. Commun. 40, 1982.
71. E.P. Ippen et.al. "Additive pulse mode locking" J. Opt. Soc. Am. B, 6 (1989).
72. J. Mark et. al. "Femtosecond pulse generation in a laser with a nonlinear external resonator." Opt. Lett. 14, 1989.
73. F. Ouellette and M. Piche, "Pulse shaping and passive modelocking with a nonlinear Michelson interferometer." 60, (1986).
74. M. Hofer et.al. "Modelocking with cross-phase and self-phase modulation." Opt. Lett. 16, (1991).
75. M. E. Fermann, et.al.,"Nonlinear amplifying loop mirror." Opt. Lett. 15, (1990).
76. K. Shtyrkova, et.al. Integrated artificial saturable absorber based on Kerr nonlinearity in silicon nitride," CLEO AF1B.6 (2017).
77. C. R. Doerr, et.al. "Additive-pulse limiting" Opt. Lett. 19, 1994.
78. C. Honninger et. al. "Q-switching stability limits of continuous-wave passive mode locking" J. Opt. Soc. Am. B 16 (1999).
79. K. J. Blow et.al. "Experimental demonstration of optical soliton switching in an all-fiber nonlinear Sagnac interferometer" Opt. Lett. 14, (1989).